\documentclass[12pt,preprint]{aastex}
\usepackage{natbib}
\usepackage{amsmath}

\newcommand{\thetahat}{$\Hat{\theta}_{mag}$}

\newcommand{\psidelay}{$\psi_{delay}$}
\newcommand{\psio}{$\psi_o$}
\newcommand{\phio}{$\phi_o$}
\newcommand{\chio}{$\chi_o$}

\newcommand{\Vvector}{$\Vec{v}$}
\newcommand{\aodpm}{$a_{0_{DPM}}$}
\newcommand{\adpm}{$a_{1_{DPM}}$}
\newcommand{\avpm}{$a_{0_{VPM}}$}

\newcommand{\el}{$\ell$}

\newcommand{\Ptime}{$P_{time}$}

\newcommand{\al}{$\alpha$}
\newcommand{\bt}{$\beta$}
\newcommand{\Pone}{$P_1$}
\newcommand{\Ptwo}{$P_2$}
\newcommand{\Pthree}{$P_3$}
\newcommand{\Pdot}{\textit{\.{P}}}

\newcommand{\chisqr}{$\chi^2$}
\newcommand{\dchisqr}{$\Delta \chi^2$}
\newcommand{\rmunits}{rad~m$^{-2}$}

\shorttitle{Quantitative Model of PSR B0809+74}
\shortauthors{Rosen and Clemens}

\begin{document}

\title{A Quantitative Model for Drifting Subpulses in PSR B0809+74}

\author{R. Rosen}
\affil{NRAO, P.O. Box 2, Green Bank, WV 24944}
\email{rrosen@nrao.edu}
\author{P. Demorest}
\affil{NRAO}

\begin{abstract} 

In this paper we analyze high time resolution single pulse data of PSR B0809+74 at 820 MHz.  We compare the subpulse phase behavior, undocumented at 820 MHz, with previously published results.  The subpulse period changes over time and we measure a subpulse phase jump, when visible, that ranges from $95^\circ$ to $147^\circ$.  We find a correlation between the subpulse modulation, subpulse phase, and orthogonal polarization modes.  This variety of complicated behavior is not well understood and is not easily explained within the framework of existing models, most of which are founded on the drifting spark model of \citet{rud75}.  We quantitatively fit our data with a non-radial oscillation model \citep{cle08} and show that the model can accurately reproduce the drifting subpulses, orthogonal polarization modes, subpulse phase jump, and can explain the correlation between all these features.

\end{abstract}

\section{Introduction}
\label{intro}

\begin{figure}
\begin{center}
\includegraphics[scale=0.7]{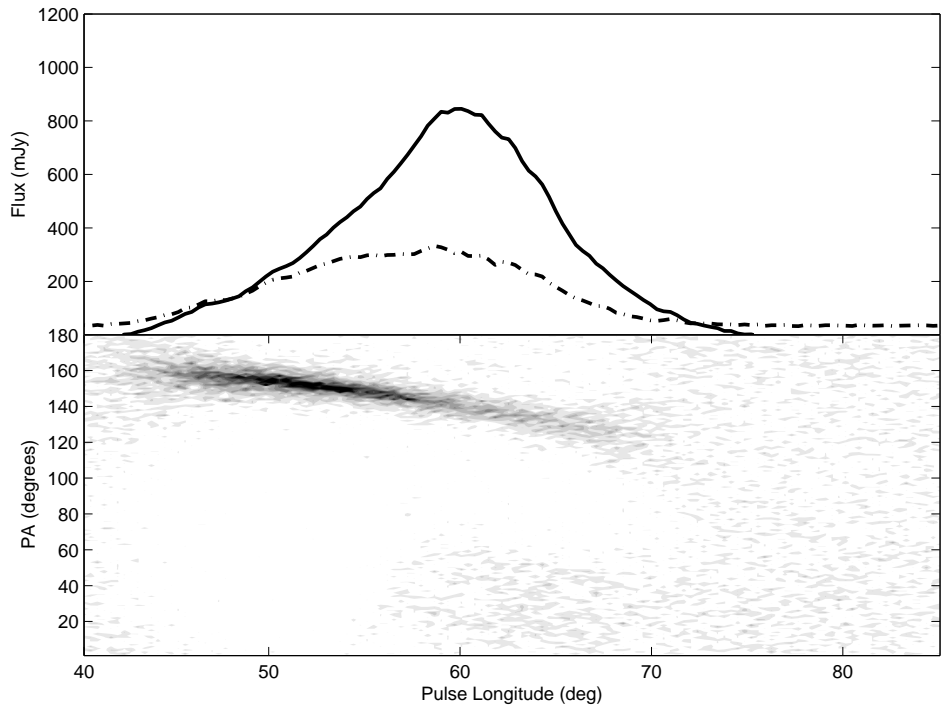}
\end{center}
\caption{Top:  The average total intensity (solid line) and linear polarization (dashed line) of PSR B0809+74 from epoch MJD 54922 at 820 MHz, consisting of 232 pulses.  Bottom:  A 2D histogram of the polarization angle for the same data.  The center of the pulse profile is arbitrary; we chose maximum to be $60^\circ$ in pulse longitude to directly compare our data to that of \citet{edw03} (see Figure \ref{fig:edw03}). }
\label{fig:AverageProfile}
\end{figure}

First discovered in 1968 by \citet{col68}, PSR B0809+74 is a bright, slow pulsar with drifting subpulses that has been continuously studied over the past 40 years.  The literature contains a wide range of behavior including changes in subpulse period, subpulse phase behavior, average pulse shape, and orthogonal polarization modes as a function of radio frequency.  The average pulse profile and polarization angle histogram for our observations at 820 MHz are shown in Figure \ref{fig:AverageProfile}.  \citet{hob04} have monitored this pulsar for at least 6 years, measuring a spin period (\Pone) of 1.292 seconds and a dispersion measure of $6.116$ pc cm$^{-3}$.  A list of basic parameters are in Table \ref{table:BasicParams}.

\begin{table}
\begin{tabular}{c|ccc}
Property & Value & Uncertainty \\
\hline
PSRJ & J0814+7429 & ... \\
RAJ & 08:14:59.50 & 0.02 s \\
DECJ & +74:29:05.70 & 0.11 '' \\
Period (\Pone) s& $1.292241446862$ & $3*10^{-12}$ \\
Dispersion Measure (DM) pm cm$^{-3}$ & $6.116$ & $0.018$ \\
Epoch & $49162.0$ & ... \\
Spin Down (\Pdot) s s$^{-1}$ & $1.68114\times10^{-16}$ & $1.4\times10^{-20}$ \\
\end{tabular}
\caption{The basic parameters of PSR B0809+74 \citep{hob04}.}
\label{table:BasicParams}
\end{table}

At low frequencies, 81.5 to 151 MHz, the measured subpulse period is around 53 ms \citep{bar81,dav84}.   The measured subpulse period is the spacing between two adjacent subpulses in the same pulse and is usually referred to as \Ptwo.  As discussed in \citet{cle04}, the value of \Ptwo~ is not an accurate measurement of the underlying fundamental subpulse period, \Ptime.  At higher frequencies, the \Ptwo~ appears to decrease to 39 ms, 31 ms, and 29 ms at 406 MHz, 1412 MHz, and 1720 MHz respectively \citep{dav84,bar81}.  Our measurements at 820 MHz fall in the middle range of observational frequencies.  \citet{bar81} find that while the subpulse period appears to change between 102.5 and 1720 MHz by a factor of 1.8, the time it takes for a subpulse to return to the same longitude, \Pthree, remains constant.

\begin{figure}
\begin{center}
\includegraphics[scale=0.6]{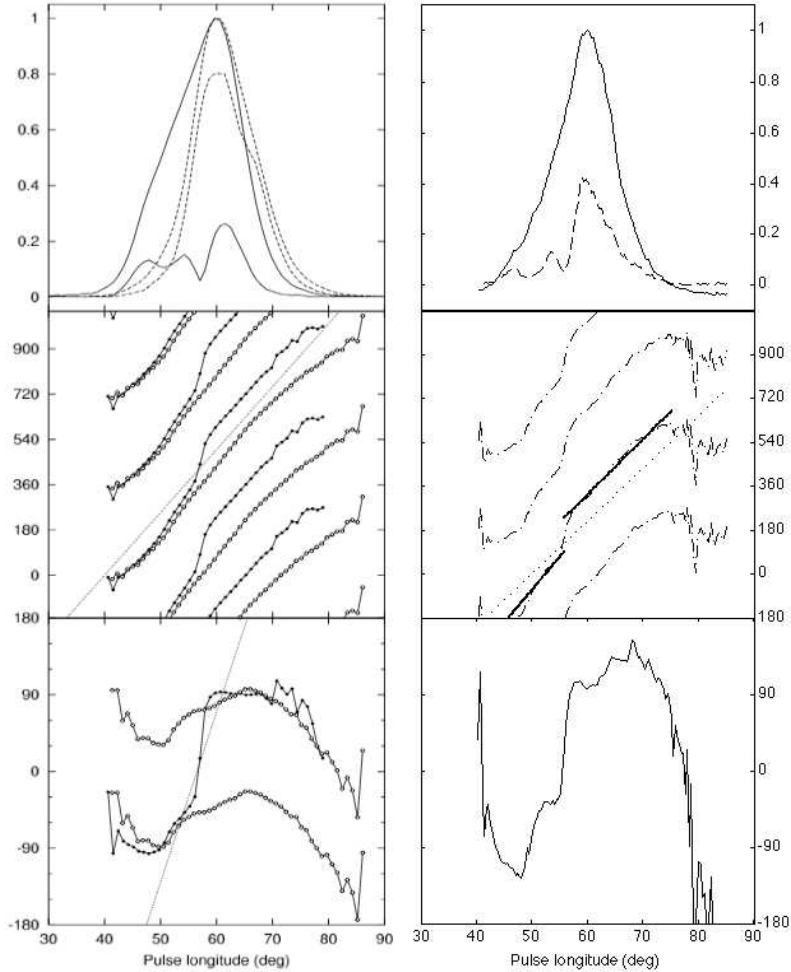}
\end{center}
\caption{Top left:  The average profile and subpulse longitude envelope for PSR B0809+74 at 328 MHz (dashed lines) and 1380 MHz (solid lines).  Top right: The average profile and subpulse longitude envelope at 820 MHz taken on MJD 54922.  Middle left:  The subpulse phase, plotted multiple times, spaced $360^\circ$ apart.  The white and dark circles are from data collected at 328 MHz and 1380 MHz, respectively.  The dotted line shows the phase slope of $25^\circ$.  Middle right:  The subpulse phase for our data at 820 MHz; the dotted line shows the phase slope of  $27.8^\circ$.  Bottom left:
The difference between the subpulse phase and that of the phase slope, indicating the magnitude of the phase jump.  The phase jump at 1380 MHz (dark circles) is $120^\circ$ and the phase jump at 328 MHz (white circles) is plotted twice with an offset of $120^\circ$.  Bottom right: The difference between the subpulse phase and that of the phase slope for our data at 820 MHz, results in a phase jump of $116^\circ$.  Using an alternative method of fitting two lines on either side of the phase jump (middle, right panel) rather than subtracting the phase slope results in a phase jump of $145^\circ$ (see \S\ref{modes}).  The phase jump at 820 MHz occurs at the same pulse longitude as the phase jump at 328 MHz and 1380 MHz.  All the plots in the left panels are reproduced from \citet{edw03}.}
\label{fig:edw03}
\end{figure}

The likely underlying cause for the change in the measured subpulse period is a subpulse phase discontinuity (or jump) that appears at high frequencies but not at low frequencies.  The subpulse phase jump is not seen at 328 MHz \citep{edw03,edw04}, 408 MHz \citep{pro86}, or at 500 MHz \citep{wol81}.  At 1380 MHz, \citet{edw03,edw04} report a phase jump of $\sim120^\circ$ as shown in the middle panel of Figure \ref{fig:edw03}.   Furthermode, the subpulse phase jump occurs starts at approximately $56.5^\circ$ in pulse longitude (bottom left panel of Figure \ref{fig:edw03}), corresponding with a minimum the subpulse amplitude envelope (dark line in the top left panel of Figure \ref{fig:edw03}).  As \citet{edw03} discuss, in any given pulse with two subpulses, the subpulses generally lie on opposite sides of the subpulse phase jump and appear closer together in pulse phase than they normally would in the absence of the phase jump, resulting in a smaller value of \Ptwo. 

Most drifting subpulse models are based on the drifting spark model \citep{rud75} in which a vacuum gap forms between the stellar surface and co-rotating magnetosphere due to the charge depletion from the emitted particles.  To prevent the vacuum gap from growing indefinitely, sparks discharge across the vacuum gap.  These sparks are fixed  relative to each other and form a carousel that rotates around the magnetic pole at a rate incommensurate with the spin period of the star.  The drifting subpulses are the manifestation of these spark discharges.  Known as the drifting or (rotating) spark model, this model is the basis for many current models of drifitng subpulses \citep{kom70,bac76,gil00}.

In \citet{cle04,cle08}, we proposed a non-radial oscillation model based on asteroseismological techniques \citep{dzi77} (see also \citet{rob82,cle00}) as an alternative to the drifting spark model.  Pulsations in stars are not uncommon: white dwarf stars, ZZ Ceti stars, rapidly oscillating AP stars, and delta Scuti stars all show pulsation modes \citep{kle98,vank00,win81,kur82,bre69}.  We were not the first proponents of a oscillation model for pulsars; \citet{gol68,vanh80,str92} all proposed oscillations as an explanation for drifting subpulses.  However, these papers did not address the wide range of phenomenology seen in pulsars with drifting subpulses.

In this paper, we analyze high quality single pulse measurements of PSR B0809+74 at 820 MHz.  In \S\ref{obs} we discuss our observations and conduct a detailed analysis of the data in \S\ref{data}.  We then explain our model in \S\ref{model} and examine the data in the context of our model in \S\ref{fit}.    Finally, in \S\ref{conc}, we discuss how our 820 MHz observations compare the observations at other frequencies and explain the single pulse behavior within the context of a non-radial oscillation model.

\section{Observations}
\label{obs}

The observations of PSR B0809+74 were taken in the spring of 2009 with the 100-m Green Bank Telescope (GBT) using the new pulsar backend GUPPI in filterbank mode.  The dates and lengths of the 10 observations are listed in Table \ref{tab:Nulls}.  Full-Stokes spectra were acquired in a 200~MHz-wide band centered at 820~MHz radio frequency.  The frequency resolution was $\sim$98~kHz and the spectrum integration time was 160~$\mu$s.  The filterbank data were then averaged into 1024-bin
single-pulse profiles using the ephemeris given in Table~\ref{table:BasicParams}.

Flux and polarization calibration were performed using the {\sc psrchive} software package \citep{hot04}.  Each of the 10 observations was performed at a different hour angle.  The rotation of the source
with respect to the telescope was used to solve for the receiver system's intrinsic polarization cross-coupling matrix, following \citet{vans04}.  From the calibrated profiles, we determined a rotation
measure (RM) of $-12.2 \pm 0.2$~\rmunits, consistent with the catalogued value of $-11.7 \pm 1.3$~\rmunits~ \citep{man72}.  The calibrated single-pulse profiles were then RM-corrected and integrated over the full band for the analysis described in the following sections.

\section{Data Analysis of PSR B0809+74}
\label{data} 

\subsection{Subpulse Period}
\label{period}

The ten observations we acquired in April 2009 varied in length and morphology, as shown in Table \ref{tab:Nulls}.  Eight of the ten observations displayed nulling, periods with zero emission.  Only two epochs, MJDs 54922 and 54944b, exhibited a single subpulse period in the FFT of the entire run; the remaining observations had multiple significant periods.  The first and last observations (MJDs 54922 and 54961) had the brightest flux.  When displaying data and our fits to the data, we use epoch MJD 54922 as it has the second largest flux and is not affected by nulls as is data from MJD 54961. 

\begin{table}
\begin{tabular}{c|ccccc}
Date & Epoch & Length of Observations & Length of Nulls & Multiple Periods & Mean Flux\\
  ... & MJD & Pulses & Pulses & ... & mJy \\
\hline
04-01-2009 & 54922 & 232 & 2, 2 & no & 310.3\\
04-22-2009 & 54943 & 466 & 0 & yes & 91.9\\
04-23-2009 & 54944a & 397 & 3 & yes & 191.9\\
04-23-2009 & 54944b & 476 & 2 & no & 189.9\\
04-27-2009 & 54948a & 475 & 3, 2, 2 & yes & 126.5\\
04-27-2009 & 54948b & 475 & 3 & yes & 171.7\\
04-27-2009 & 54948c & 476 & 2, 2, 2, 2 & ?? & 114.3\\
04-27-2009 & 54948d & 476 & 0 & yes & 60.2\\
04-27-2009 & 54948e & 477& 3 & yes & 40.3\\
05-10-2009 & 54961 & 475 & 8, 4 & yes & 374.6\\
\end{tabular}
\caption{A summary of our observations at 820 MHz.  The first and last observations (MJDs 54922 and 54961) have the brightest flux.  When displaying data and our fits to the data, we use epoch MJD 54922 as it has the second largest flux and does not display nulls.}
\label{tab:Nulls}
\end{table}

We find the fundamental subpulse period, \Ptime, to be between 48 and 54 ms, which is consistent with the measured value of \Ptwo~ by \citet{dav84,bar81} at low frequencies without the presence of a subpulse phase jump (see \S\ref{params}).  A fast Fourier transform (FFT) shows that the subpulse alias peaks at approximately 39 ms.  However, the FFT calculates the period of the subpulse based on the separation of the peaks, and the presence of the subpulse phase jump makes the subpulse peaks artificially closer than the would be in the absence of the phase jump \citep{edw03}.  

The average pulse profile at 820 MHz is consistent with those at 328 MHz and 1380 MHz.  As shown in the top panels of Figure \ref{fig:edw03}, the subpulse modulation envelope at 820 MHz (dashed line, top right panel) resembles that at 1380 MHz but not at 328 MHz (top left panel).  This difference is the result of at subpulse phase jump (shown in the bottom panels of Figure \ref{fig:edw03}) at 820 MHz and 1380 MHz; we discuss the subpulse phase behavior in \S\ref{modes}. 

\begin{figure}
\begin{center}
\includegraphics[scale=.8]{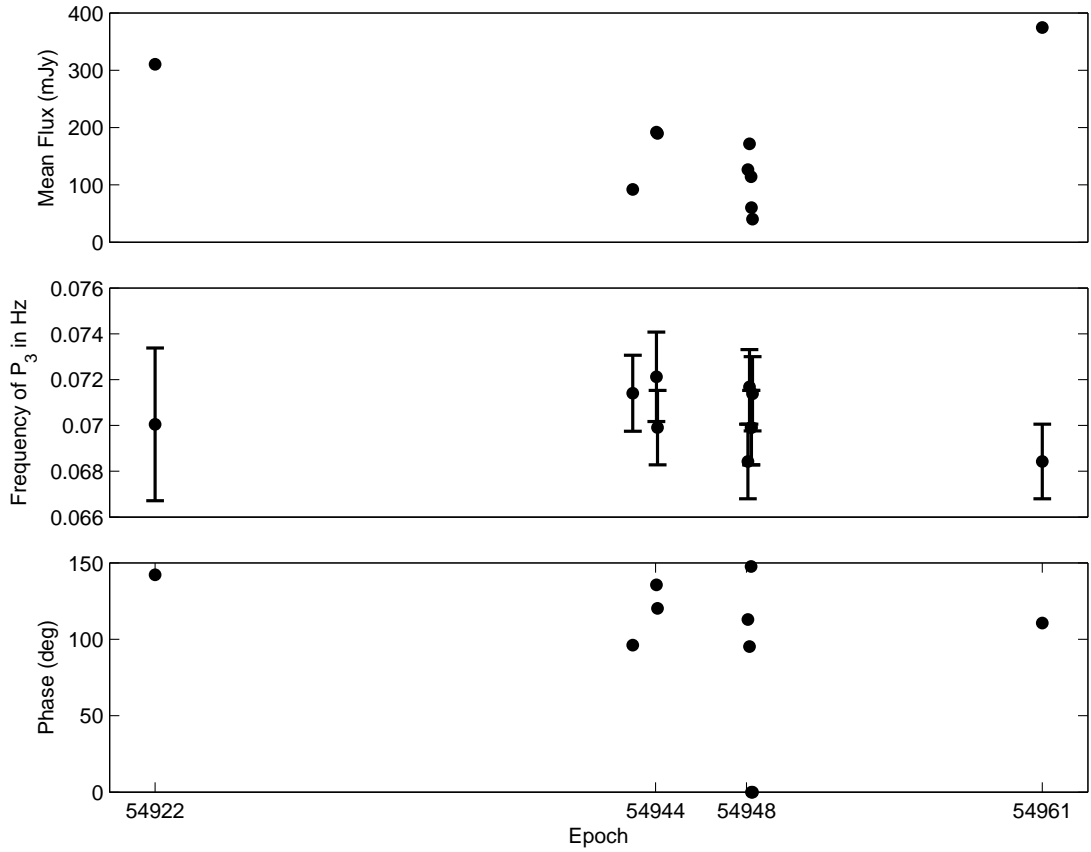}
\end{center}
\caption{Top panel:  The mean flux as a function of epoch.  Middle panel:  The period \Pthree~ at each epoch.  Bottom panel:  The subpulse phase jump; see \S\ref{modes} for our methodology on calculating the phase jump.}
\label{fig:AllMJDs}
\end{figure}

Figure \ref{fig:AllMJDs} characterizes the subpulse behavior for each observation.  The mean flux in the top panel of Figure \ref{fig:AllMJDs} is also described in Table \ref{tab:Nulls}.  The subpulse period can also be characterized by \Pthree, which is the time it takes for a subpulse to return to the same longitude after successive spins of the star.  \Pthree~ is not sensitive to which alias we chose from the FFT as the subpulse period.  To measure \Pthree, we use the longitude resolved fluctuation spectrum (LRFS), a Fourier transform calculated at each spin phase.  The value of \Pthree~ for each observation is shown in the middle panel in Figure \ref{fig:AllMJDs}.  The bottom panel displays the subpulse phase jump; the method for calculating the subpulse phase jump is discussed in \S\ref{modes}.

\begin{figure}
\begin{center}
\includegraphics[scale=.8]{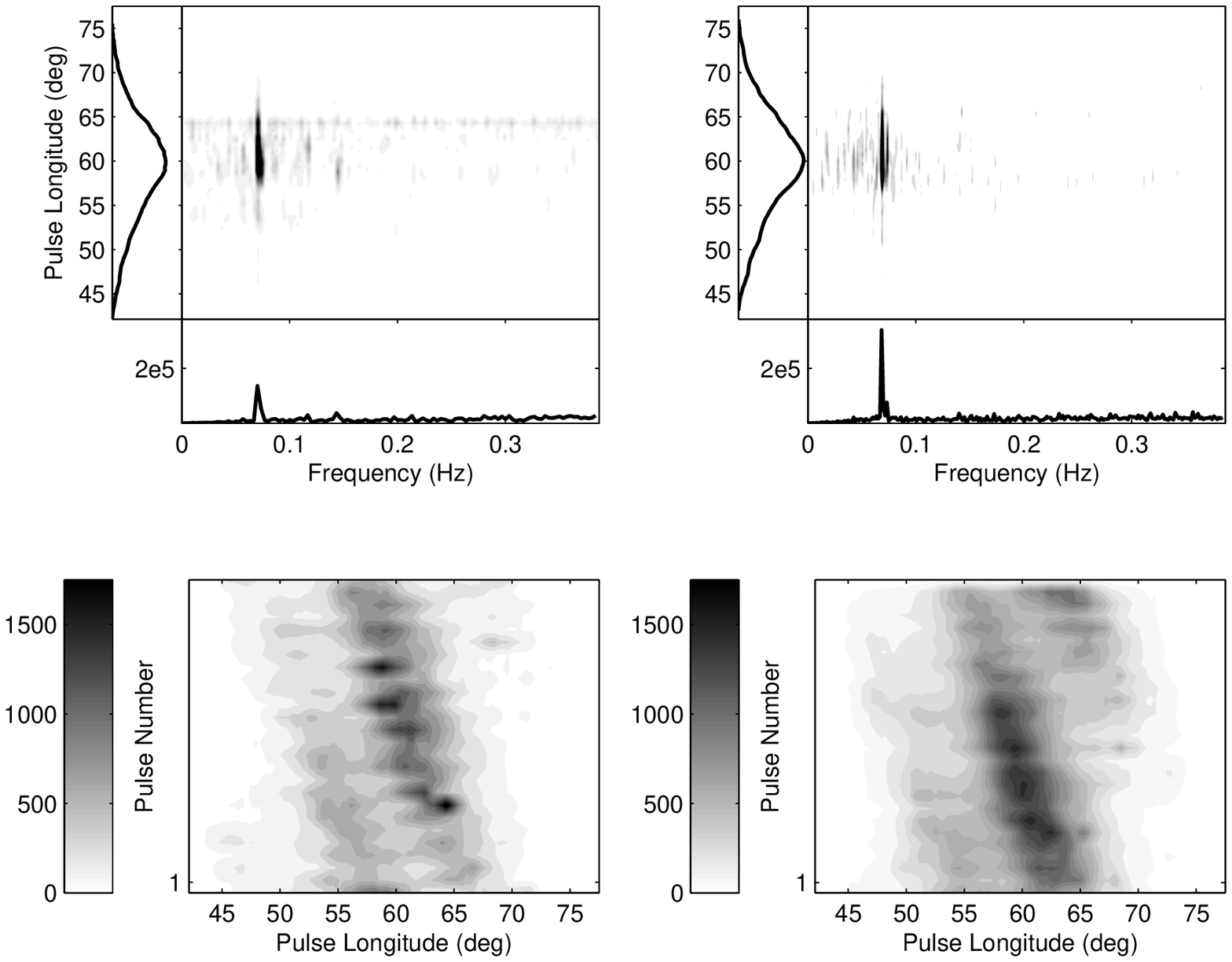}
\end{center} 
\caption{Top: The longitude resolved fluctuation spectrum from MJDs 54922 (left) and 54961 (right), where \Pthree~ $\sim$11.31\Pone.  The left panels show the integrated pulse profile; the maximum on the plot is 1100 mJy.  The bottom panels show the power in mJy$^2$/Hz. Bottom:  The same data set folded at \Pthree.}
\label{fig:LRFS}
\end{figure}

The top plots in Figure \ref{fig:LRFS} shows the LRFS of two different data sets taken at different epochs; the bottom panels of the top plots show a peak at \Pthree~ = 14.61 seconds, or $\sim$11.31~\Pone.  The bottom plots of Figure \ref{fig:LRFS} shows the corresponding \textit{driftband} plot, a contour plot of the data folded at \Pthree.  Figure \ref{fig:AllMJDs} shows \Pthree, calculated from the LRFS, for all the different epochs.

\subsection{Subpulse Phase and Orthogonal Polarization modes}
\label{modes}

Each of the ten data sets show a subpulse phase jump at pulse longitude of $56.5^\circ$ (offset by $3.5^\circ$ from the maximum in the pulse profile), as shown in the right panels of Figure \ref{fig:edw03}.   We calculate the subpulse phase using the amplitude and phase of the peak of \Pthree~ in the LRFS.  Figure \ref{fig:edw03} shows the subpulse modulation envelope, the subpulse phase, and the subpulse phase with the nominal phase slope of $28.7^\circ / ^\circ$ removed (bottom right panel).  Instead of fitting a line to the subpulse phase jump like \citet{edw03}, we fit two lines: one on each side of the phase jump.  The solid lines in the right third panel of Figure \ref{fig:edw03} illustrates this fitting.  The difference between the two parallel solid lines at pulse longitude $56.5^\circ$ is the subpulse phase jump; in Figure \ref{fig:edw03}, the jump is $145.1^\circ$.  The middle panel of Figure \ref{fig:AllMJDs} shows the subpulse phase jump for each epoch and the bottom panel shows the average slope (taken from the two lines fit on either side of the jump).  The magnitude of the jump, present in 7 of the 10 data sets, ranges from $95^\circ$ to $147^\circ$.

The orthogonal polarization modes have distinctly different behavior on the leading and trailing edge of the pulse profile; the behavior changes at $56.5^\circ$ as well, coincident with the subpulse phase jump.  As shown in Figure \ref{fig:AverageProfile}, the left side of the pulse profile appears to be dominated by a single polarization mode; the right side of the pulse profile shows a combination of two orthogonal polarization modes.  This is consistent with the polarization behavior described by \citet{edw04} at 1380 MHz.

\section{Our Non-Radial Oscillation Model}
\label{model}

In previous papers, we developed an oblique pulsator model \citep{kur82} for pulsars in which drifting subpulses are produced by non-radial oscillations whose periods are incommensurate with the spin period of the pulsar \citep{cle04,cle08}. The non-radial modes of our model are aligned to the magnetic axis, so in addition to the drifting time-like pulses, our model produces longitude stationary variations caused by nodal lines rotating past our line of sight \citep{cle04}.  Nodal lines are places of unmodulated emission and are described by the zeros in the spherical harmonic in Equation \ref{eqn:dpm}.  The emission on either side of a nodal line is out-of-phase and the subpulse period is defined as rate at which a region between two nodal lines emits radio emission.

\begin{figure}
\begin{center}
\includegraphics[scale=0.4]{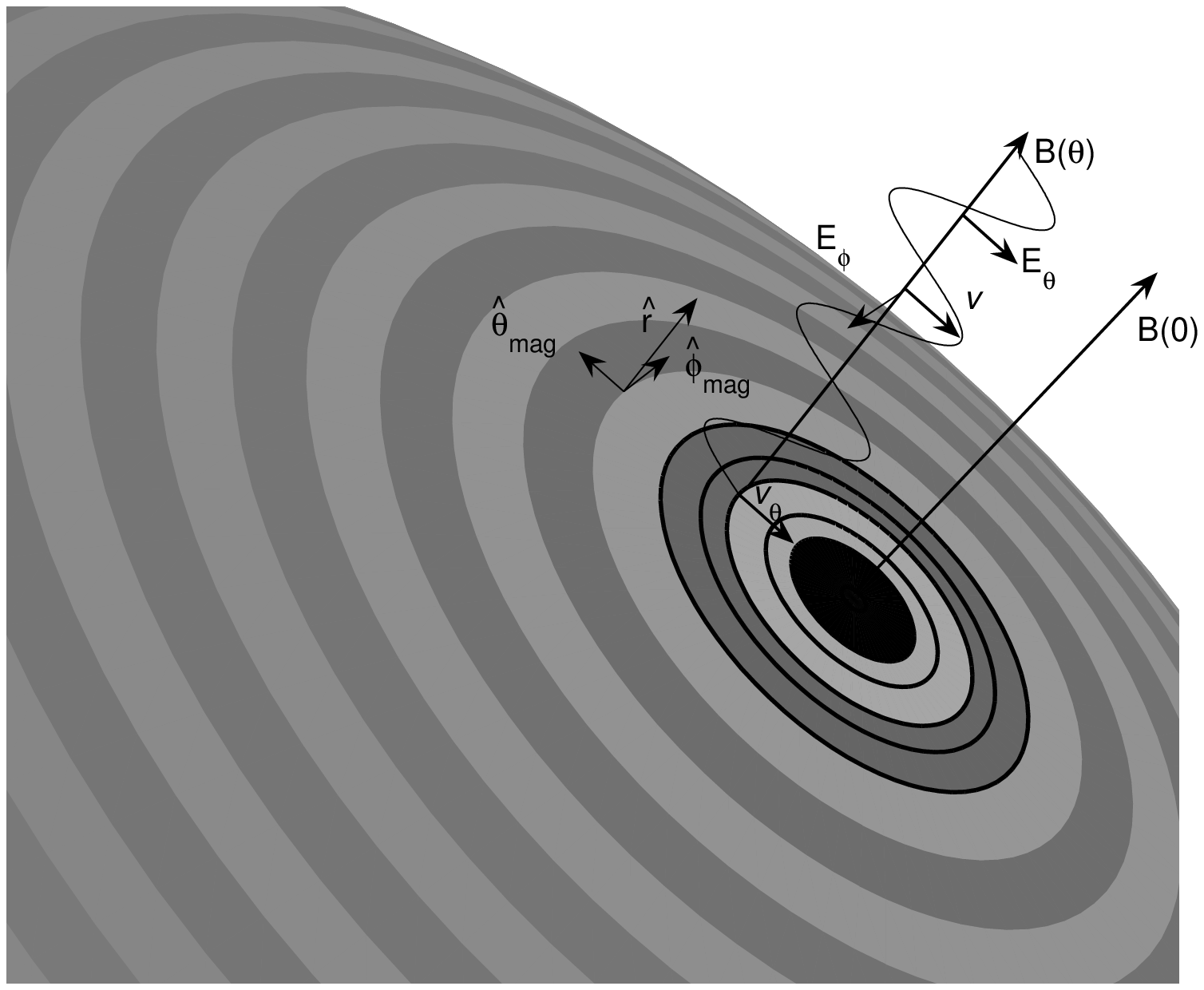}
\end{center}
\caption{The polarization geometry near the surface of a neutron star.  The magnetic field $B_{star}$ extends outward, normal to the stellar surface.  The electric field has two components: $E_{\Hat{\theta}}$ points in a longitudinal direction and $E_{\Hat{\phi}}$ is oriented in a latitudinal direction with respect to the magnetic pole.  The dominant velocity vector \Vvector~ points in the $\pm$ \thetahat~ direction, toward and away from the magnetic pole.}
\label{fig:dipole}
\end{figure}

The pulsations cause displacements of stellar material which modulate the linearly-polarized emission, as might be produced by curvature radiation.  These displacements have a transverse electric field vector that points toward the magnetic pole and follows the rotating vector model of \citet{rad69} (see Figure \ref{fig:dipole}).  This mode of polarization, the \textit{displacement polarization mode}, can be described as \citep{cle08}:

\begin{equation}
\label{eqn:dpm}
A_{DPM}(t) = a_{0_{DPM}} + a_{1_{DPM}} \Psi_{l,m=0}(\theta_{mag})\\\cos(\omega{t}-\psi_0-\psi_{delay}))
\end{equation}
\\where $\Psi_{l,m=0}$ is a spherical harmonic of high \el~ and $m=0$.  The variable $\theta_{mag}$ refers to the magnetic co-latitude, because the pulsations in our model are aligned to the magnetic pole.  The subpulse period, $\omega$, is related to the fundamental subpulse period, \Ptime = $2\pi/\omega$.  The phase term \psio~ allows for the arbitrary phase of the drifting subpulses.  The \psidelay~ term allows for a time lag between the maximum amplitude of the pulsations and emission maximum; for non-adiabatic oscillations, the thermal maximum can lag the displacements in phase and \psidelay~ allows for this effect \citep{cle08}.   The amplitudes $a_{0_{DPM}}$ and $a_{1_{DPM}}$ are to be fitted to the data, as well as \Ptime, \psio, and \psidelay.   

The pulsational displacements and their associated velocities move in the plane of the magnetic field.  Thus, the induced electric field as a result of the velocities ($\vec{E} = \vec{v}\times\vec{B}$) is naturally orthogonal to the Radhakrishnan and Cooke vector (see Figure \ref{fig:dipole}).  The polarization mode that results from the induced electric field due to the velocities is the \textit{velocity polarization mode} and is expressed as:

\begin{equation}
\label{eqn:vpm}
A_{VPM}(t) = a_{0_{VPM}}{{\frac{\partial{\Psi_{l,m=0}}}{\partial{\theta_{mag}}}}}\sin(\omega{t} - \psi_0),
\end{equation}
\\which incorporates the time derivative and the $\theta_{mag}$ derivative of Equation \ref{eqn:dpm}, as appropriate for horizontal pulsation velocities.  This equation is analogous to the $V_{\theta}$ in equation three of \citet{dzi77}. 

Thermal and field emission from the neutron star surface can accelerate electrons along open field lines with the formation of a vacuum gap \citep{jes01}.  \citet{str92} proposed that neutron star oscillations could modulate the radio intensity if greater quantities of plasma are injected into the magnetosphere during pulsation maxima, when local heating of the stellar surface is greatest.  As discussed in \citet{cle08}, we assume that the amplitude of the displacement polarization mode follows surface thermal variations caused by non-radial oscillations.  This means that for non-adiabatic oscillations, the thermal maximum can lag the displacements in phase, parameterized by \psidelay~ in Equation \ref{eqn:dpm}. 
 
The parameters in our model can be divided into two groups: geometrical parameters and pulsational parameters \citep{ros08}.   The geometrical parameters, discussed in \S\ref{geometry}, are largely independent; only \bt, the angle between the magnetic pole and our line-of-sight, and \el, the degree of the spherical harmonic, are related.  There are 7 pulsational parameters - amplitudes, periods, and phases and all of these are fit within our model.

\section{Quantitative Fitting of PSR B0809+74}
\label{fit}

The fitting process occurs in two steps: we fit the polarization angle swing to determine the pulsar geometry and then we fit the data to our non-radial oscillation model to determine the pulsational parameters.  Since we do not expect the geometry of the star to change over time, we added the polarization angle from each of the data sets together determined the geometry from a single fit.  Because our model does not incorporate circular polarization, we removed the Stokes paramter $V$ from the total intensity such that $I_{new} = \sqrt{I^2-V^2}$.  

\subsection{Pulsar Geometry}
\label{geometry}

We fit the polarization angle swing to determine the pulsar geometry, namely \al~ (the offset in rotation and magnetic axes), \bt~ (the angle between the magnetic pole and our line of sight), \phio~ (the rotational longitude of the magnetic pole),  and \chio~ (the position angle of the linear polarization at \phio).  The only pulsational parameter that interacts with these parameters is the spherical harmonic degree \el~ because the positions of the nodal lines that encircle the magnetic pole are related \el, and the path our sightline crosses through these nodal lines is related to \bt.

\begin{figure}
\begin{center}
\includegraphics[scale=.8]{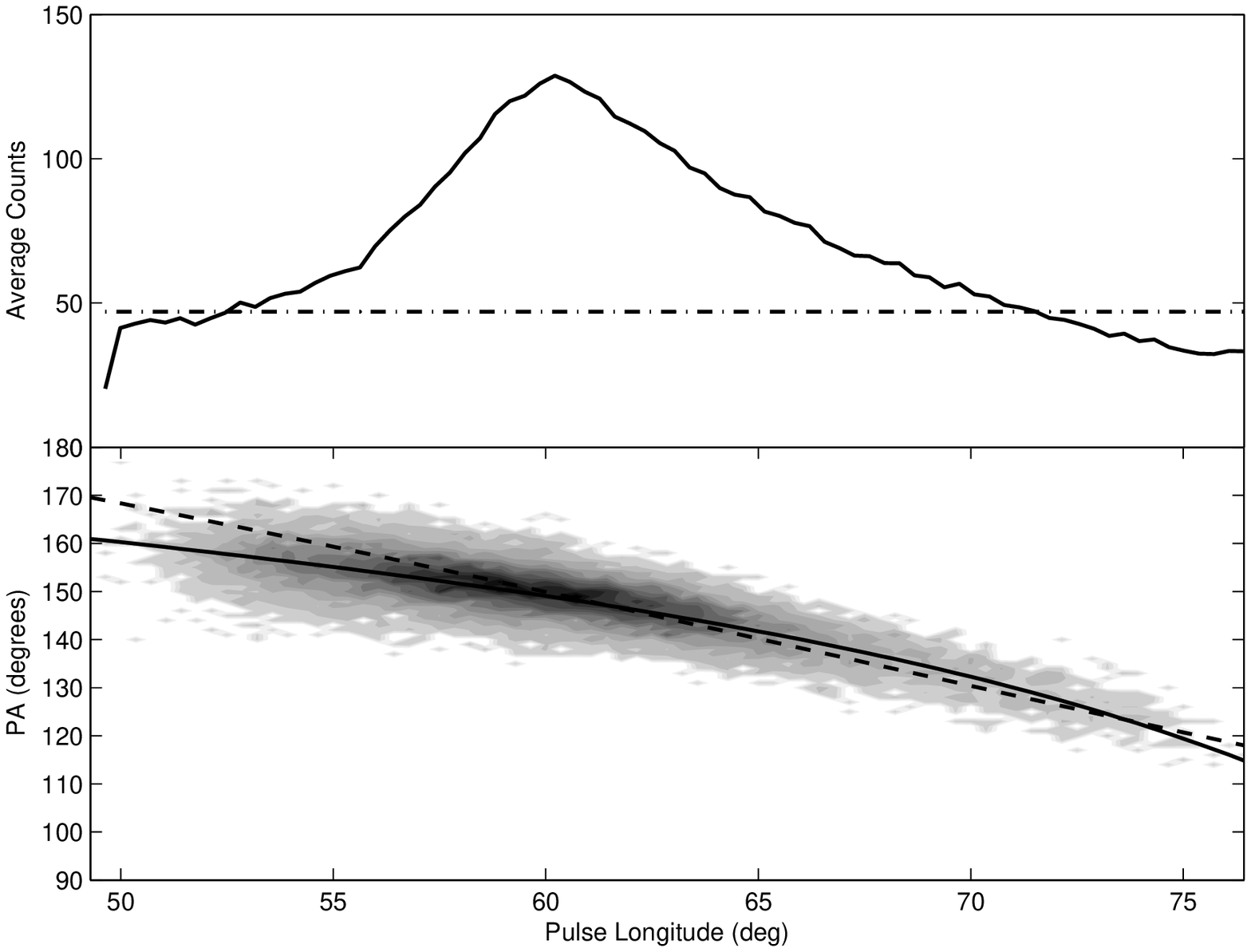}
\end{center}
\caption{Top panel:  The average number of counts per bin for each polarization angle in the histogram.  The dotted line represents the $2\sigma$ threshold; polarization angles below this threshold are not fit.  Bottom panel:  The fitted portion of the polarization angle histogram (above the threshold) using the combined data from all ten data sets.  The solid black line is the result of our fitting process; the dashed line is using published values (see text).}
\label{fig:AngleFits}
\end{figure}

To determine the geometry, we use the polarization angle histogram rather than the individual polarization angle measurements \citep{eve01}.  To do this, we compute a single histogram of the polarization angle using all ten data sets.  We then discard all polarization angles less than $90^\circ$, so that we fit a single polarization mode.  Since the polarization angle is a histogram, each bin has a distinct number of counts.  Using the counts in each bin (not counting the discarded polarization angles less than $90^\circ$), we calculated the standard deviation for the counts per bin.  We fit only the polarization angle bins in the histogram that have counts greater than $2\sigma$.  The top panel of Figure \ref{fig:AngleFits} shows the average number of counts per bin for each polarization angle.  The dotted line represents our $2\sigma$ cut off; we only fit values above this line.  This process prevents our fits from being dominated by the noise.

To fit our data, we convert each count in each bin in the polarization angle histogram above the $2\sigma$ threshold to x-y values.  If a bin at a given polarization angle has 20 counts, we create 20 x-y pairs at that value.  The more counts per bin, the more x-y pairs are created at the polarization angle.  We do this for all the bins above the threshold and then fit all the x-y points.

\begin{figure}
\begin{center}
\includegraphics[scale=.8]{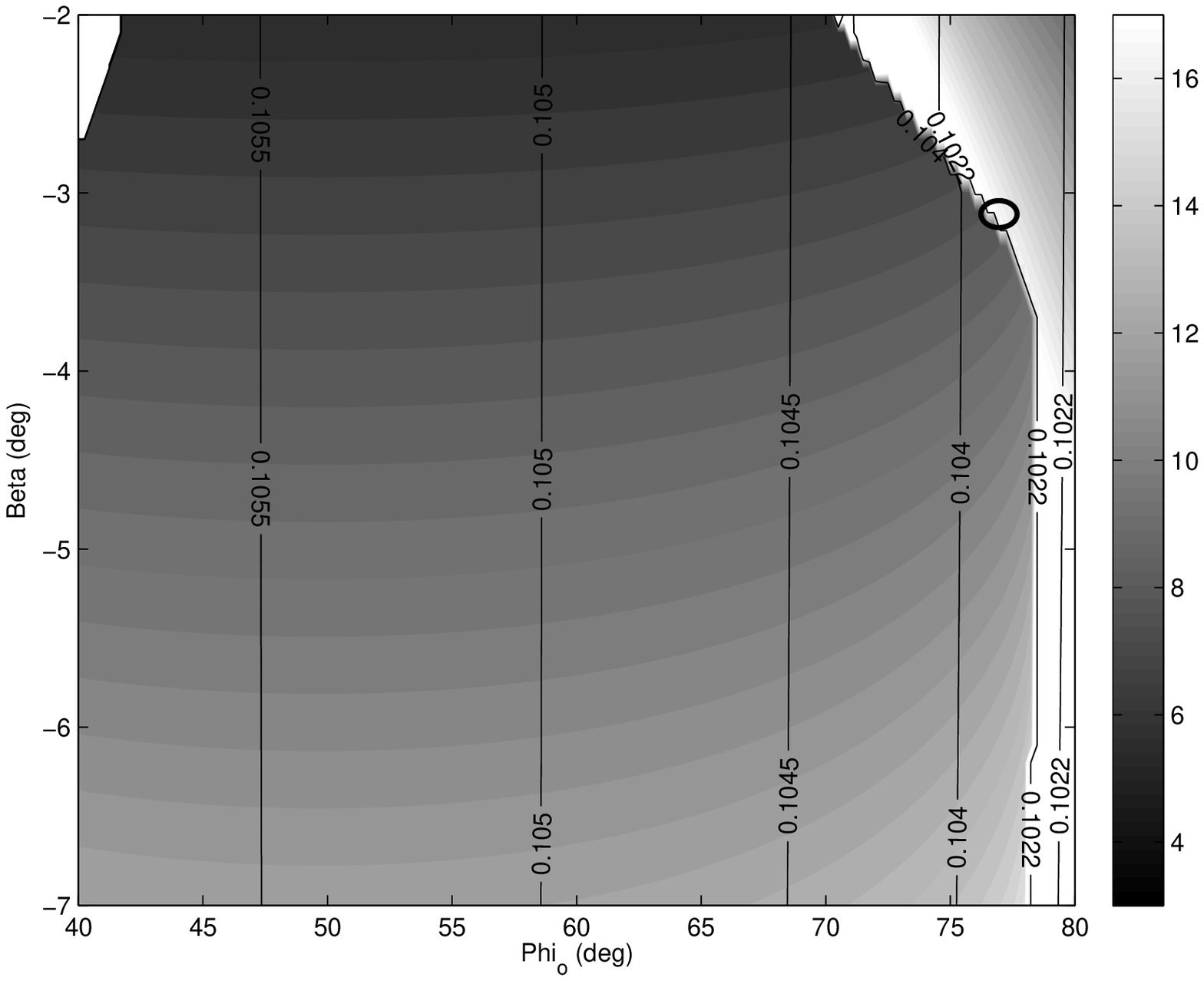}
\end{center}
\caption{The grayscale contours show \al~ for each of the values of \bt~ and \phio.  The solid lines trace the standard error.  We used increments of 0.1 and 0.25 degrees for \bt~ and \phio~ respectively.  The black circle marks the values of \bt~ abd \phio~ which result in the smallest standard error.}
\label{fig:AllAngle_map}
\end{figure}

In \citet{ros08}, we found that the four geometrical parameters (\al, \bt, \chio, \phio) cannot be fit independently and it is the ratio of \al~ to \bt~ that is significant; for any value of \al~ and \chio, corresponding values of \bt~ and \phio~ could be found with approximately the same goodness of fit.  Therefore, to determine the geometry, we fit \al~ and \chio~ for various values of \bt~ and \phio.  Figure \ref{fig:AllAngle_map} shows a map of the standard error for all trial values of \bt~ and \phio.  We determined the best geometry from the combination of these two parameters that has the smallest standard error, denoted by the circle in Figure \ref{fig:AllAngle_map}.

We find that a \bt~ of $-3.1^\circ$ and a \phio~ of $76.75^\circ$ had the smallest standard error.  For these values of \bt~ and \phio~, the best fit results in an \al~ of $16.81^\circ$ and a \chio~ of $-199.29^\circ$.  Interestingly, the fit places the closest crossing of the magnetic pole, \phio, on the trailing edge of the pulse profile, offset from the maximum intensity.  This is similar B0656+14: the average polarization angle shows a shallow slope slightly more curved at the tail end of the pulse profile \citep{eve01}.  \citet{eve01} fit the polarization angle and find \phio~ to be offset from the maximum in the total intensity by $14.9^\circ$ in pulse longitude; in our case, \phio~ is offset from the maximum in the total intensity by $19.17^\circ$.  

These values differ from that of \citet{ran93b}, who found \al~ and \bt~ to be $9^\circ$ and $4.5^\circ$, respectively.  \citet{ran93} determines the geometry based on the half-power width of the pulse profiles and the beam radius at 1 GHz.  The difference between the values of \citet{ran93b} and our fits is most likely due the different approach to determining the geometry and to the change in geometry with respect to observing frequency \citep{smi06}.  We find that the value of \phio, while at the trailing edge of the pulse profile, is consistent with the standard polar cap size.  Using similar analysis to that of \citet{wel09}, we find that the emission height is about $0.08R_{LC}$, where $R_{LC}$ is the light cylinder radius, and that the size of the magnetic cap is between the magnetic axis and the last open field line.  Even if we assume that the steepest polarization angle swing coincides with the center of the pulse profile and half of the emission region is missing, the polar cap size does not extend past the last open field line. 

In the bottom panel of Figure \ref{fig:AngleFits}, we show the portion of the histogram data that we use to fit polarization angle as well as our fit.  We also show the polarization angle using the values in \citet{ran93b} and assuming \phio~ is at the maximum of the pulse profile ($\sim18^\circ$ in Figure \ref{fig:AllAngle_map} or $56.5^\circ$ in pulse longitude) and \chio~ = $-134^\circ$.  However, for the fit using the values in \citet{ran93b}, we flipped the sign on \bt~ for internal consistency \citep{eve01}.


\subsection{Pulsational Parameters}
\label{params}

To verify that a longer subpulse period, consistent with the literature at low frequencies, matches the data better than a period of 39 ms, we fit our model to the data which we discuss in detail in \S\ref{params}.  For each of the 10 data sets, we fit the full range of observed subpulse aliases from 31.5 to 51.5 ms.  For 9 out of 10 scans, a smaller root mean square (RMS) residual is obtained for a subpulse period in the range 48 to 54 ms than the 39 ms where the power in the FFT peaks.  We fix the subpulse period in our fits rather than letting it be a free parameter because several data sets show that the peak due to the subpulse period has a secondary peak of lesser amplitude at a spacing several time larger than the resolution in the FFT.  We speculate that the presence of nulls in some of the data sets causes the subpulse period to wander throughout the observations (see Table \ref{tab:Nulls}).  \citet{vanl03} show similar behavior where the subpulse drift rate in PSR B0809+74 changes after nulling.

After establishing the pulsar geometry, we fit the pulsational parameters, namely: the spherical degree, \el, the amplitudes of the displacement and velocity polarization modes (\adpm, \aodpm, and \avpm), the arbitrary phase of the drifting subpulses (\psio), and the phase that allows for a time lag between the maximum amplitude of the pulsations and emission maximum (\psidelay).  Because the subpulse period has has a small secondary peak in some of the data sets (see \S\ref{period}), we fix the subpulse period based on the subpulse alias in the FFT (see below).  The subpulse period is similar but unique for each data set.

As with our fits to the polarization angle, we fit each data set to our model for various values of \el.  For each data set, the value of \el~ that returns the smallest \chisqr~ is the best fit to the data.  We treat each data set separately in determining the goodness of fit because of the variation of stochastic pulse amplitudes.  The data sets that have small variations in pulse amplitude will have a smaller standard errors compared to data sets that have large variations.  We estimate the noise, $\sigma_i$ in our data using the radiometer equation:

\begin{equation}
\sigma_i = \frac{T_{sys}}{\sqrt{B\tau}}
\label{eqn:rad}
\end{equation}
\\where $T_{sys}$ is the system equivalent flux density (SEFD) in mJy ($\sim 14-18$ Jy depending on the epoch), $B$ is the bandwidth (200 MHz), and $\tau$ is the integration time (160 $\mu$s).

\begin{figure}
\begin{center}
\includegraphics[scale=.7]{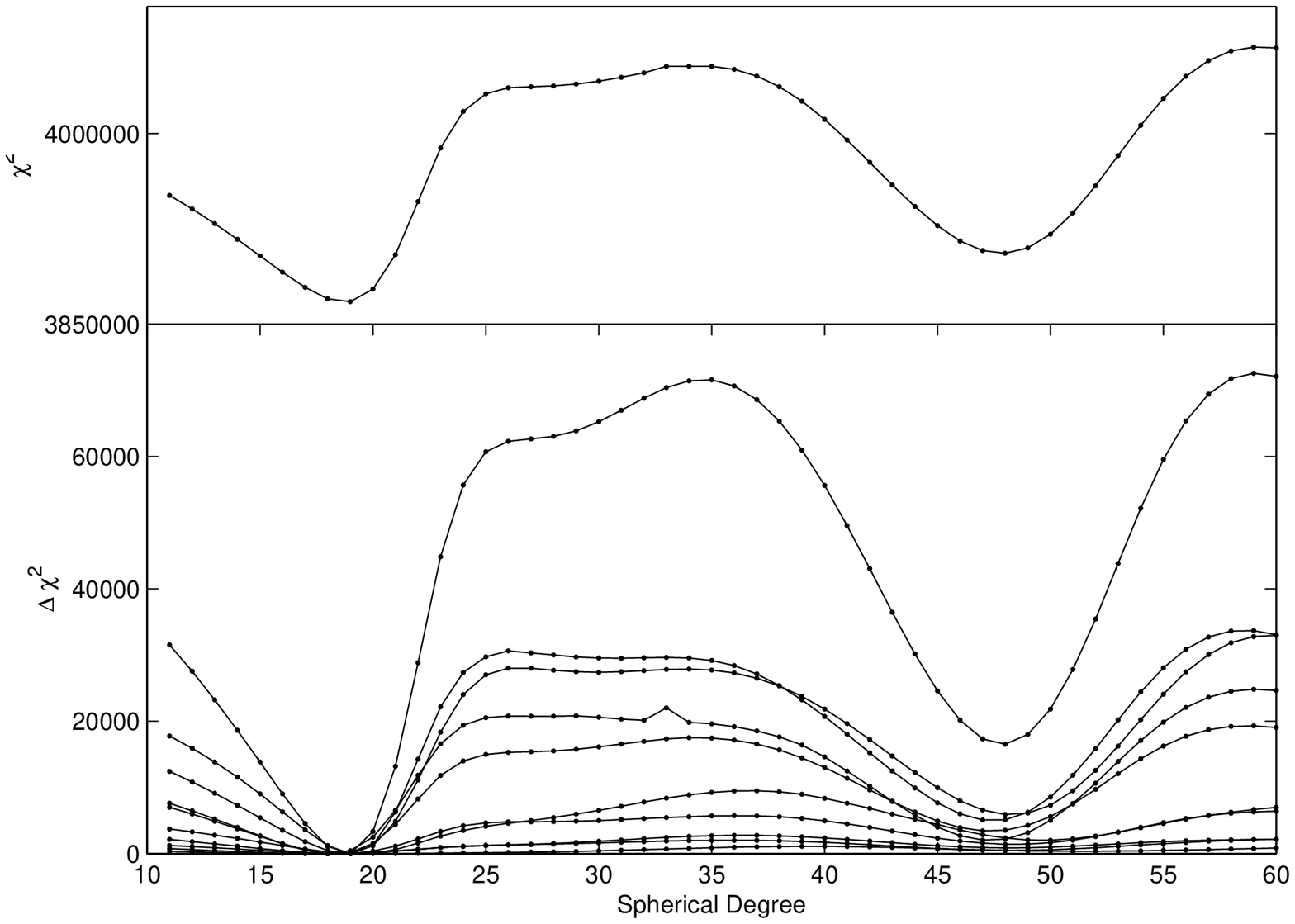}
\end{center}
\caption{Top: The total plot of the \chisqr~ for various values of \el, in increments of one.   Bottom: A plot of the \dchisqr~ for each epoch.}
\label{fig:l_map}
\end{figure}

The top panel Figure \ref{fig:l_map} shows total \chisqr~ for all the data; the bottom panel shows the individual \dchisqr~  for each epoch as a function of spherical degree, \el.  Each dot represents one trial, integer value of \el.  For all data sets, \el~ is either 18 or 19; the difference is within the error of our fit.  The best value of \el~ based on the total \chisqr~ is 19.

\begin{figure}
\begin{center}
\includegraphics[scale=.7]{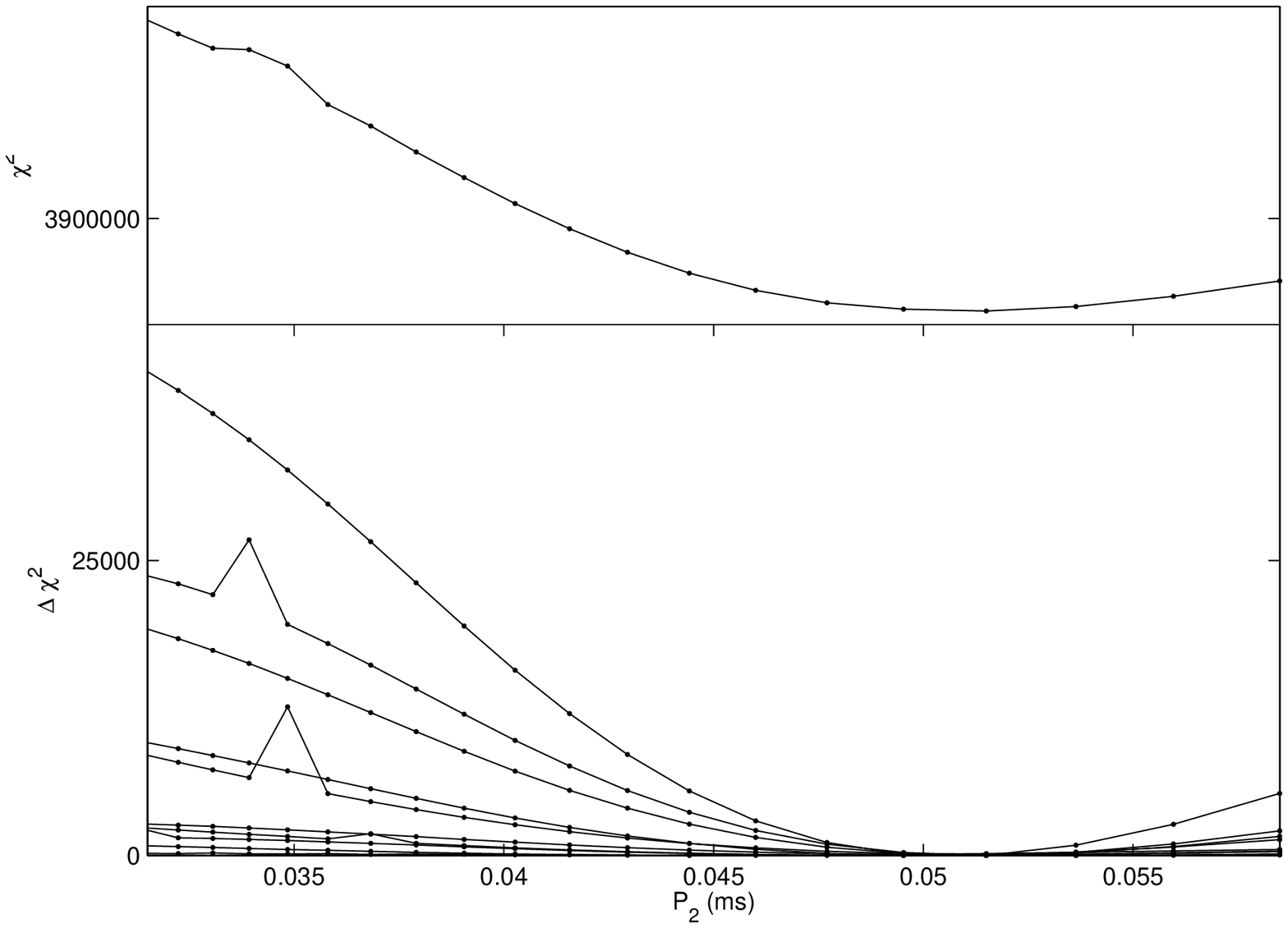}
\end{center}
\caption{The total plot of the \chisqr~ for various values of the subpulse period.  Each dot represents one alias of the subpulse period taken from the FFT.  We fit subpulse aliases from 31.5 to 58.5 ms.  Bottom: A plot of the \dchisqr~  for each epoch.}
\label{fig:subpulse_map}
\end{figure}

We use a method similar to that for finding the best value of \el~ to find the best subpulse period.  Since our model fits the data in real space ($I$, $Q$, and $U$) rather than Fourier space, and a subpulse phase jump is a natural part of the model due to the presence of a nodal line, fitting the data to our model is more accurate method for determining the subpulse period than using the FFT.  For each data set, we choose a subpulse alias from the FFT and fit our model to the data, calculating \chisqr.  We repeat this for every subpulse alias from 31.5 to 58.5 ms for each data set, using the value of the subpulse period with the smallest \chisqr~ as the best fit.  The top panel of Figure \ref{fig:subpulse_map} shows the total \chisqr~ for values of the subpulse period using different subpulse aliases in the FFT from 31.5 to 51.5 ms; each dot is one alias from the FFT.  The bottom panel shows the individual \dchisqr~ for each epoch.  The subpulse period for all data sets except for one is between 48 and 54 ms; the best subpulse period based on the total \chisqr~ is 51.5 ms. 

\begin{figure}
\begin{center}
\includegraphics[scale=.7]{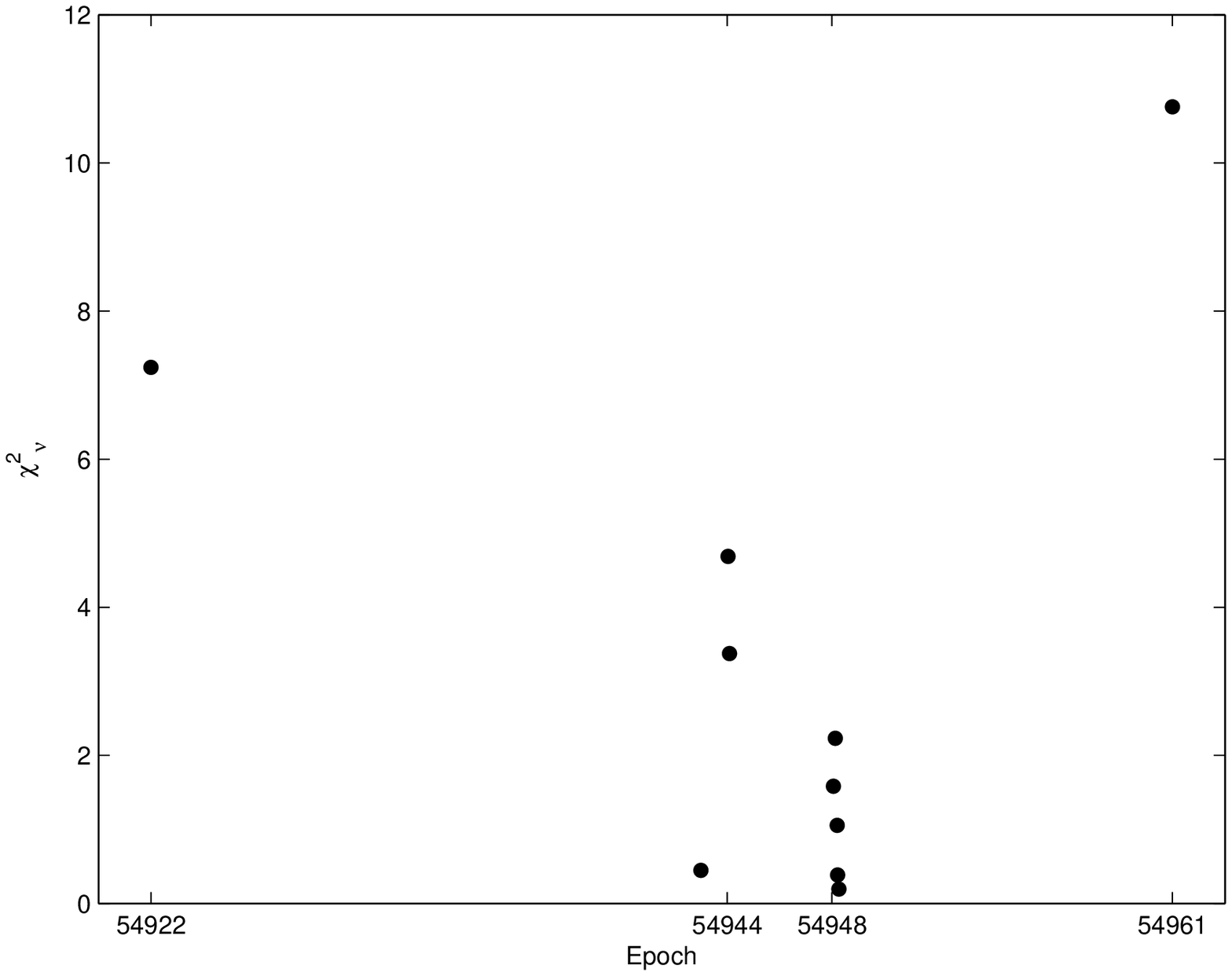}
\end{center}
\caption{We fit the data to our model, where the amplitudes and phases are the free parameters.  We set the values of \el~ and the subpulse period based on the methodology described in the text.  The plot shows reduced \chisqr~ (${\chi^2}_\nu$) resulting from out fit for each epoch.}
\label{fig:chisqr_v}
\end{figure}

Once we determine the best value of \el~ and the subpulse period for each data set, we use Equations \ref{eqn:dpm} and \ref{eqn:vpm} to fit our single pulse data to the model.  From Equations \ref{eqn:dpm} and \ref{eqn:vpm}, we can create the primed Stokes parameters, the Stokes parameters in the non-rotating frame of the star \citep{ros08}.  However, instead of defining an emission window as we did in \citet{ros08}, we multiply the Stokes parameter $I$ by the average pulse shape.  We then used equations 7, 8, and 9 from \citet{ros08} to transform the primed Stokes parameters into the observer's frame of reference (unprimed space), and similarly multiplied $L$, which is invariant under this transformation, by the average linear polarization.  For each epoch, we use the best value of \el~ and the subpulse period, as determined in the process outlined above, and fit the remaining parameters (\adpm, \aodpm, \avpm, \psio, and\psidelay).  Figure \ref{fig:chisqr_v} shows the reduced \chisqr~ (${\chi^2}_\nu$) for each epoch.  The value of ${\chi^2}_\nu$ strongly correlates with the mean flux (see the top panel of Figure \ref{fig:AllMJDs}).  Table \ref{table:params} lists the best fit for the geometrical and pulsational parameters for MJD 54922.   

\begin{table}
\begin{tabular}{c|c}
Parameter & Value \\
\hline
$\alpha$ & $16.81^\circ$\\
$\beta$ & $ -3.1^\circ$\\
$\phi_o$ & $76.75^\circ$\\
$\chi_o$ & $-199.29^\circ$\\
$a_{0_{DPM}}$ & 1.201\\
$a_{1_{DPM}}$ & 0.733\\
$a_{0_{VPM}}$ & 0.055\\
$\psi_0$ & $-85.66^\circ$\\
$\psi_{delay} $ & $113.857^\circ$\\
$P_2$ & 51.503 ms\\
$l$ & 19\\
${\chi^2}_\nu$ & 7.24\\
\end{tabular}
\caption{The geometrical and pulsational parameters resulting from our fit of the data from 54922 to the model.}
\label{table:params}
\end{table}

\begin{figure}
\begin{center}
\includegraphics[scale=.8]{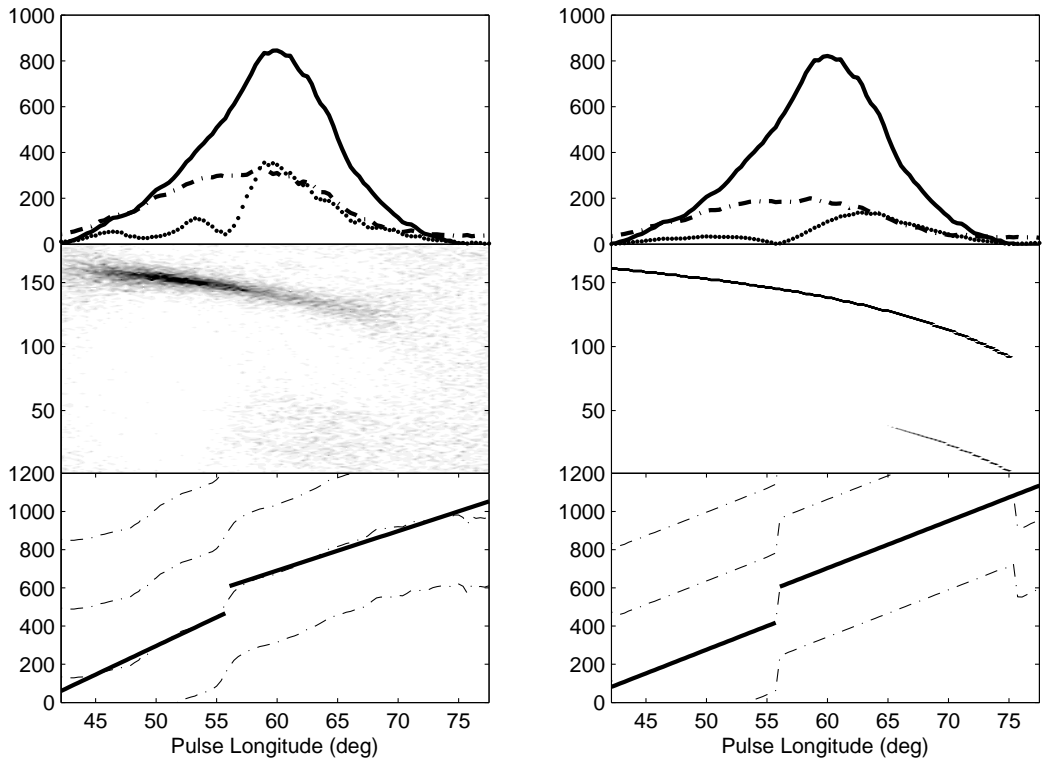}
\end{center}
\caption{Top panels: The average pulse profile (solid line), the average linear polarization (dashed-dotted line), and the subpulse modulation envelope (dotted line) of the data from MJD 54922 (left) and model (right).  Middle panels:  The polarization angle histogram for both the data and model.  Bottom panel:  The subpulse phase, calculated from the LRFS (see \S\ref{modes}).  The solid line is a linear fit to both components of the subpulse phase.}
\label{fig:AverageProfile_sim}
\end{figure}

\begin{figure}
\begin{center}
\includegraphics[scale=.8]{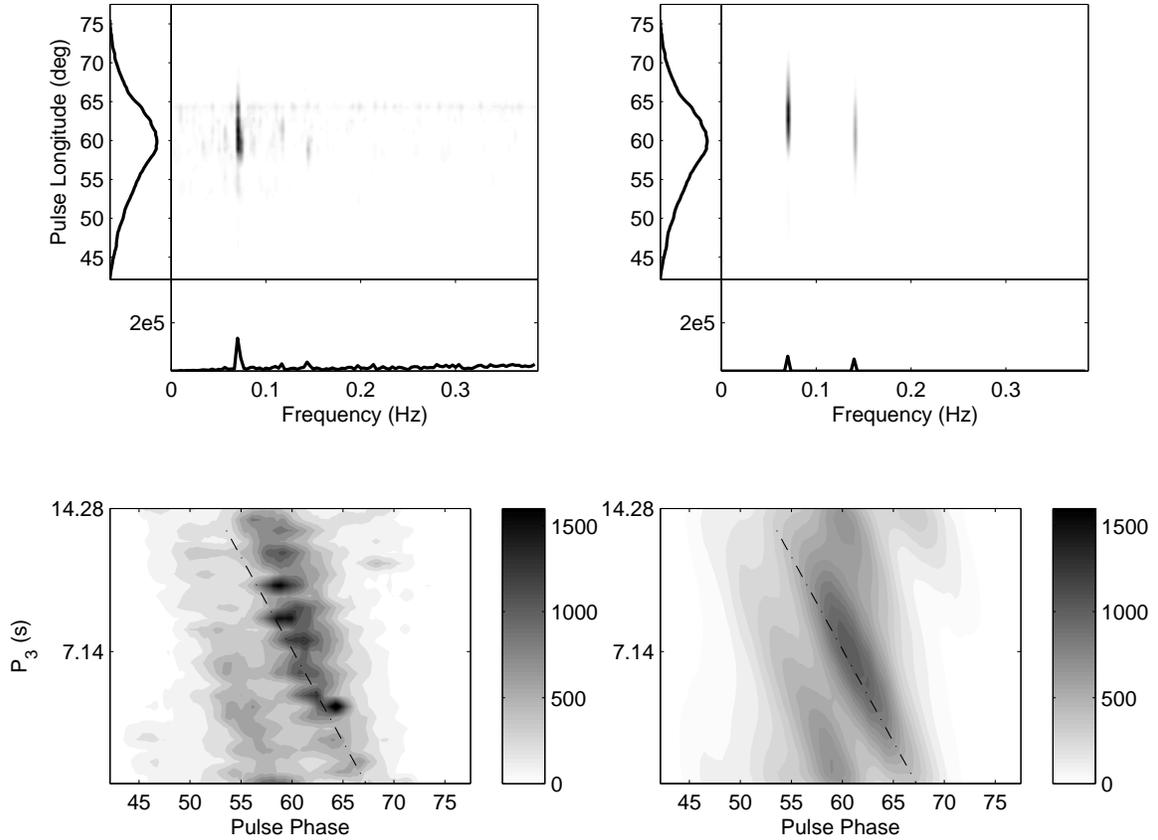}
\end{center}
\caption{Left column: the LRFS (top) and driftband of data from MJD 54922. The left panel shows the integrated pulse profile; the maximum on the plot is 1100 mJy.  The bottom panel shows the power in mJy$^2$/Hz.  Right column: the LRFS and driftband of simulated data using the best parameters as determined by our fit of the data to our model.  The panels in the LRFS have the same scale and units as the data in the top left plot.}
\label{fig:LRFS_sim}
\end{figure}

\begin{figure}
\begin{center}
\includegraphics[scale=.8]{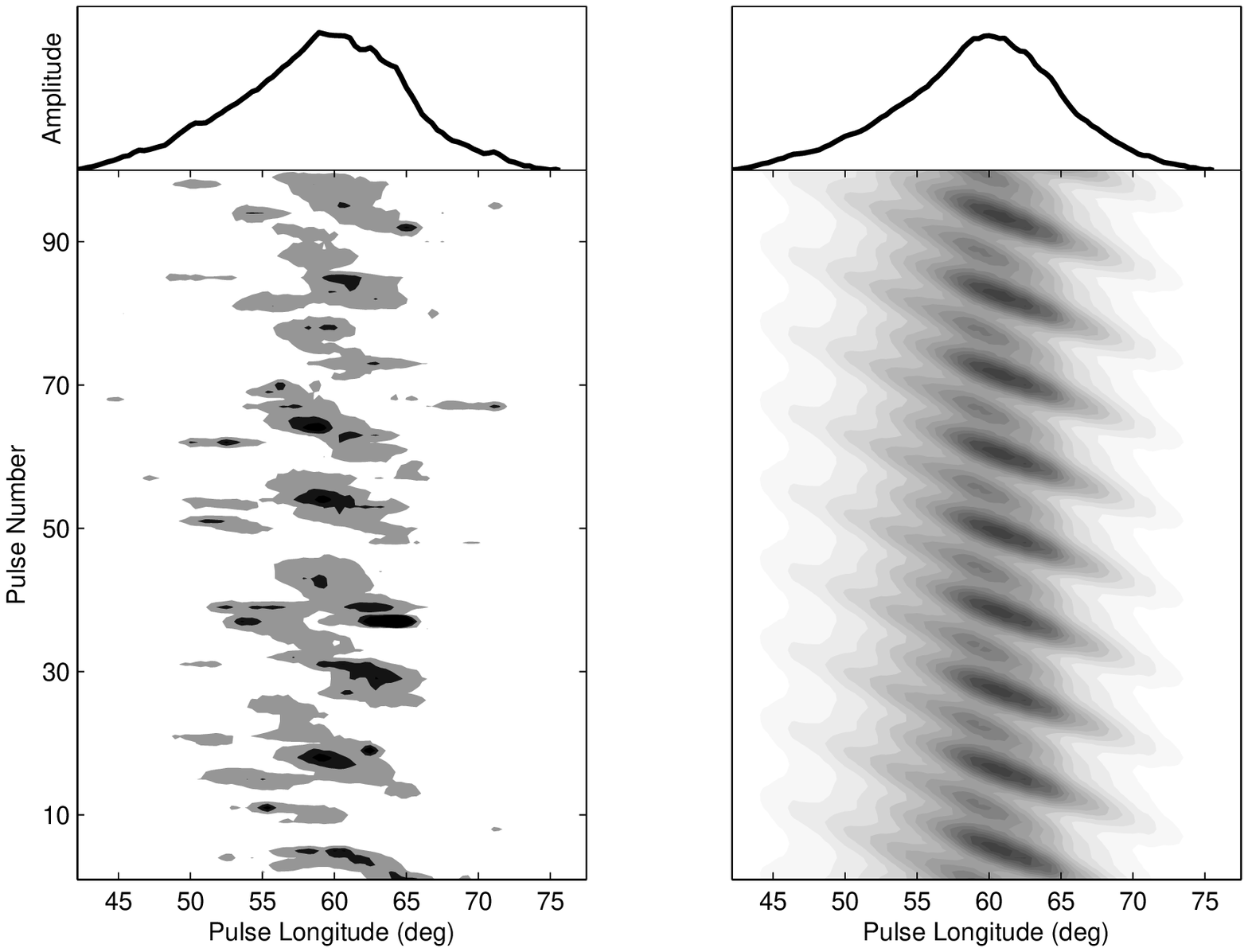}
\end{center}
\caption{Left panels:  A sequence of 100 pulses from data from MJD 54922 (lower panel) and the average profile (upper panel).  Right panels:  A sequence of 100 pulses created from the fitted parameters from our model against the data from this epoch.}
\label{fig:singlepulses}
\end{figure}

Using the fitted parameters, we create synthetic lightcurves and compare them to the data.   Figure \ref{fig:AverageProfile_sim} shows data from MJD 54922 and a simulation using our fitted parameters.  The data shows a phase jump of $145.1^\circ$ while the data produced by our model has a phase jump of $187.7^\circ$.  The fact that our model does not reproduce the phase jump exactly is not surprising; the phase jump is fit indirectly in our model.  Our model fits $I$, $Q$, and $U$ directly and the phase is computed secondarily.  The phase jump occurs at $56.5^\circ$, coincident with the nodal line, as a result of our best fit of \el.  Figure \ref{fig:LRFS_sim} shows the longitude resolved fluctuation spectrum and driftbands of the data from MJD 54922 and simulation.   The harmonic in the simulation of 0.14 Hz is more pronounced than that of the data.  This is mainly due to the fact that the data neither contains pure sinusoids nor a perfect window function, while our simulation does.  The two phase terms in our fitting, \psio~ and \psidelay, are most prominent in the driftband plot.  In the data, the maximum intensity occurs between a pulse phase of $60^\circ$ and $65^\circ$; altering \psio~ changes the location of the maximum intensity in both pulse phase and \Pthree.  In the simulation, the driftband has a feature at a pulse phase of $65^\circ$ and period of 14 seconds.  This small feature, less obvious in the data, is due to the velocity polarization mode.  The value of \psidelay~ dictates the spacing of between the two polarization modes.  Figure \ref{fig:singlepulses} shows single pulses of the same data and simulations as in Figure \ref{fig:AverageProfile_sim}.  In the simulations, the subpulses on the leading edge of the profile are due to the velocity polarization mode and are slightly weaker than those due to the displacement polarization mode.

\section{Discussion}
\label{conc}

\subsection{Observations In the Context of a Non-radial Oscillation Model}

A non-radial oscillation model explains the wide range of behaviors seen in slow pulsars and is substantially simpler and more cohesive model than those based on drifting sparks.  We expect that the subpulse period should not change with radio frequency.  We also expect that a nodal line must correspond with a region of zero modulated emission and a subpulse phase jump.  

The radio frequency dependence of \Ptwo~ is a matter of debate.  \citet{dav84} found that for PSR B0809+74, the subpulse spacing (\Ptwo) appears to change with frequency, but \Pthree~ does not.  \citet{izv93} found similar results for four additional pulsars.  However, the average pulse shape also changes with frequency \citep{izv93}, resulting from the divergence of the magnetic field and the change in radial distance from the magnetic pole.  High frequencies originate close to the stellar surface, so the average pulse profile is narrow, but emission at these frequencies originates further from the magnetic pole radially than low frequencies \citep{kom70,cor78,smi06}.  

This scaling law, called radius-to-frequency mapping, can be mistakenly applied to \Ptwo~ in two ways. As the average pulse profile broadens at lower frequencies, the spacing between the subpulse increases as well \citep{rud75,izv93,wol81,bar81,gil96}.  However, as \citet{edw02} describe, the change in \Ptwo~ with frequency is a result of the change in window function: as the average pulse profile broadens (or narrows) the window function that modulates the subpulse period changes as well, affecting the apparent spacing between subpulses.   Secondly,  as illustrated in PSR B0809+74, the average pulse profile and subpulse phase behavior can change quite noticeably between different frequency.  The appearance of a subpulse phase jump will change the measured value of \Ptwo~ (see \S\ref{period}).  Both the subpulse period and phase envelope should be independent of frequency \citep{edw02,cle04}.
 
Because the oscillation mode is a fundamental property of the star, we expect the subpulse period to be independent of observing frequency.  However, the subpulse period can be distorted by both the presence of a nodal line and the width of the average profile \citep{cle04}, and therefore the apparent spacing of \Ptwo~ is not an accurate measurement of a stable, underlying clock.  In the case of PSR B0809+74, since the change in frequency effectively changes our sightline traverse, \bt~ \citep{smi06}, the subpulse phase jump that appears at higher frequencies can be attributed to a nodal line moving into our line of sight due to the change in effective geometry.  The subpulse period should remain the same, but the apparent spacing changes.  We see this in the period of PSR B0809+74: the FFT incorrectly determines subpulse spacing to be around 39 ms, but fitting the single pulses more accurately measures of the subpulse period to be 48 to 54 ms.  While our measurements of the subpulse period appear to be consistent with previous published results \citep{bar81,dav84,edw03}, the true test to determine if the subpulse period is independent of frequency is to conduct simultaneous, multifrequency observations.


The subpulse phase jump appears to change with observing frequency.  \citet{edw04, edw03} reports a phase jump of approximately $45^\circ$ and $120^\circ$ in the total intensity at 328 MHz and 1380 MHz.  Using the methodology of \citet{edw03}, we find the subpulse phase jump to be $116^\circ$ at 820 MHz; using a slightly different technique described in \S\ref{modes}, we find the phase jump to be $145^\circ$.  Regardless, the appearance of the phase jump at 820 MHz and above is consistent with a nodal line moving into our line-of-sight.  This hypothesis is borne out in the subpulse modulation envelope shown in the left upper panel of Figure \ref{fig:edw03}.  The dotted lines in the top left panel show the average profile and subpulse modulation envelope at 328 MHz; the solid lines are the same for 1380 MHz.  The subpulse modulation envelope at 328 MHz is a single peak, while the modulation envelop at 1380 MHz shows a minimum which would be consistent with the presence of a nodal line.

In our model, the choice of \el~=19 places a nodal line at $56.5^\circ$ and produces the phase jump at the same location.  The combination of a nodal line and \phio~ offset from the center of the pulse profile is responsible for the presence of two orthogonal polarization modes on the trailing edge of the pulse profile.  While the amplitude of the displacement and velocity polarization modes are constant throughout the simulation, the asymmetry due to \el~ and \phio~ cause the velocity polarization mode to be suppressed on the leading edge.

This value of \el~ is significantly smaller than the value we calculated for PSR B0943+10 (\el~=75) \citep{ros08}.  In \citet{cle04} we estimated the magnification factor to map the \el~ we observe to the surface of the star.  We found that at 1 GHz, the magnification factor was $\sim 7$, and thus in our qualitative model an \el~ of 85 translated to an \el~ of 600 on the stellar surface.  In \citet{ros08}, we found \el~ to be 75 using 430 MHz archival data \citep{des01}.  Assuming that observations at lower frequencies are at higher altitudes from the surface of the star, and thus the magnetic field is more divergent, the magnification factor is then a lower limit and the \el~ of 75 translates to a lower limit of \el~=525.  In these data, with an \el~ of 18 or 19, the lower limit becomes an \el~ of 130 on the stellar surface.    We note here that the value of \el~ = 75 for PSR B0943+10 was not rigorously determined using the methodology outlined in this paper, and smaller values for \el~ for that star are possible.

\section{Conclusions}

This paper shows the second quantitative fit of our non-radial oscillation model to single pulse data of a pulsar \citep{ros08}.  In this paper, we:

\begin{itemize}
\item show the subpulse period and subpulse phase jump, previously unpublished at 820 MHz.
\item show the subpulse phase jump and subpulse period changes with epoch.
\item are able to quantitatively determine the best value of \el~ and the subpulse period.  Our method for determining the subpulse period is more accurate than using the FFT as it determines the subpulse period in real space rather than frequency space, which can be affected by the subpulse phase jump.  Our fits using $I$, $Q$, and $U$ account for the subpulse phase jump since it is a natural part of the model.
\item are able to quantitatively fit single pulse data to our model and determine a goodness of fit using  \chisqr~ and ${\chi^2}_\nu$~ statistics.
\item can create simulations based on our fitted parameters which accurately reproduce the subpulse period, subpulse phase jump, and orthogonal polarization modes.
\end{itemize}

The morphology of PSR B0809+74 is explained easily and naturally within a pulsational model.  Our non-radial oscillation model is based on established asteroseismological principles that have explained white dwarf variations for the past 40 years \citep{dzi77}.  This model is a viable alternative to the drifting spark model and can provide physical insight in the emission mechanism and physical structure of neutron stars.


\acknowledgements
The National Radio Astronomy Observatory is a facility of the National Science Foundation operated under cooperative agreement by Associated Universities, Inc.  We would like to thank Geoffrey Wright for his thoughtful comments.

\bibliographystyle{apj}

\end{document}